# Creative Diversity: Patterns in the Creative Habits of ~10,000 People

Eric Berlow[1], Spencer Canon[2], Kaustuv DeBiswas[1], David Gurman[1,3], Shawna Jacoby[1,4], Lizbet Simmons[1], Andy Walshe[4], Rich Williams[1], Tiffany Yuan[1,4], and Mark Runco[5]

**Abstract:**
Despite popular media interest in uncovering the "creative habits" of successful people, there is a surprising paucity of empirical research on the diversity of tendencies and preferences people have when engaged in creative work. We developed a simple survey that characterized 42 creative habits along 21 independent dimensions. Data from 9,633 respondents revealed seven "Creative Species," or clusters of people with combinations of creative habits that tend to co-occur more than expected by chance. These emergent clusters where relatively stable to random subsampling of the population and to variation in model parameters. The seven Creative Species, self-sorted along a primary gradient from those characterized by more 'deliberate' creative habits (e.g., Monotasker, Risk Averse, Routine Seeker, Tenacious, Make it Happen) to those characterized by more 'open' creative habits (e.g. Multitasker, Risk Friendly, Novelty Seeker, Reframer, Let it Happen). A weaker second gradient was defined by 'outward/rational' vs 'inward/intuitive' creators. For the subset of respondents with data about their broad professional discipline (Arts/Design, Science/Engineering, and Business/Entrepreneurship) and gender, some groups were more (or less) common in some Creative Species than expected by chance, but the absolute magnitude of these differences were generally small; and knowing a person's discipline or gender was not a good single predictor of their creative preferences or tendencies. Together these results suggest that independent of discipline or gender, people vary widely in the habits, behaviors, and contexts in which they feel most creative. Understanding creative diversity is critical for improving the creative performance of both individuals and collaborative teams.

**Introduction:**

A simple internet search for "habits of creative people" returns hundreds of hits with a variation of the title, "The Top 10 Habits of Highly Creative People." One method used in these articles and essays is to pick a few famous people and attribute their creative success to their specific habits, such as their propensity to go for walks, stick to a schedule, or take risks. It is then suggested to be more creative we should do the same. These methods have two problems: First, they are logically flawed by assuming that, because this handful of elite performers have certain traits, their high performance must be attributable to those particular traits. Second, they foster the myth that there is one way to be creative. If you behave like these top performers, you'll be creative; and if you don't share their habits, you won't be. These stories conflate "how creative are you" with "how do you create".

[1] Vibrant Data Labs, Berkeley CA, USA
[2] Superotter, Bishop CA, USA
[3] KAYA Climb, San Francisco CA, USA
[4] Liminal Collective, Los Angeles CA USA
[5] Southern Oregon University, Ashland, OR USA

Despite the popular interest in the "creative habits," there is a surprising paucity of empirical research on the topic. A recent study[1] analyzed keyword themes across 1,355 creativity research studies published over 25 years (1990-2015). Of the 5,048 total keywords in the corpus, there was only one with the word "habit" ("work habits"), and it was present in only one paper.[2] Only 47 papers (3%) had keywords that include "style" (e.g., "creative style,", "cognitive style,", "learning style,", "problem solving style,", etc.). Meanwhile 15% of all published papers had at least one keyword with the word "performance," and 2 of the 3 largest keyword theme clusters, comprising 25% of the papers, were performance related. Thus, while there is an abundance of research devoted to measuring and predicting *creative performance*, there is comparatively little published work characterizing diversity among people *in how they create*. There are case studies that describe how small sets of famous creators did their work[3–6] but such efforts are difficult to generalize. There has been research on how people's cognitive tendencies and backgrounds differ between disciplines (e.g., "domain differences")[5,7–9], but the sample sizes of these studies is generally too small to evaluate the full spectrum of creative diversity within vs between domains.

Understanding how people vary in the behaviors, preferences, and contexts in which they feel the most creative – hereafter referred to as "creative habits" – is critical for any individual to better capitalize on their strengths, understand their weaknesses, and collaborate with people who are different. In this study, we addressed three broad questions: 1) Is everyone unique, or do people cluster into distinct "creative species" – or groups that share similar combinations of creative habits; 2) Do creative habits differ predictably by broad discipline (Arts vs Sciences vs Business); and 3) Do creative habits differ predictably by gender?

**Results:**

We developed, over three phases (see Methods), a survey which characterizes 42 creative habits along 21 independent dimensions. Of the 9,647 respondents that completed at least 90% of the survey, 9,633 had strong enough preferences to be assigned at least 3 creative habit tags, with 50% having at least 14 (see *Methods: Data Analysis: Assigning Creative Habits*). Non-parametric Spearman Rank Correlations between all possible pairs of ordinal questions suggest that each successfully reflected a relatively independent "creative habit dimension." The maximum rank correlation was only 0.38 and the average rank correlation across all question pairs was 0.03. Every creative habit was present in at least 10% of the population, and no single creative habit was anywhere close to ubiquitous – the most common creative habit ("novelty seeker") was present in 48% of the respondents (Figure 1a, Table 1).

The most common creative habits included the tendency to feel more creative when breaking up routines and having novel experiences, and when working on solo projects, in private spaces, without distraction (Figure 1a: "Novelty Seeker", "Hate Distractions", "Solo Creator", "Private"). The least common creative habits were the stimulation of working in a public setting, specializing on one thing, and being self-assured in one's creative work (Figure 1a: "Public", "Specialist", "Self-Assured"). However, the smooth frequency distribution of creative habits

(Figure 1a) did not show signs of any set standing out from the rest as being unusually common or uncommon relative to the rest.

A network of survey respondents was created by linking people if they shared similar sets of creative habits (see *Methods: Data analysis: Identifying "Creative Species"*). This network produced 30 clusters – or groups of people with similar combinations of habits, and the top seven clusters comprised >98% of the population (Figure 1b). Each of the seven largest clusters consistently had a high match in randomly sub-sampled networks (Supplementary Information Figure 1a, average Weighted Jaccard Similarity = 0.68). The remaining 23 small clusters, all of which contained <1% of the population, had inconsistent best matches in the sampled networks (Supplementary Information Figure 1a, average Weighted Jaccard Similarity = 0.32). These results suggest that the seven largest clusters were stable to random sub-sampling of the population. Similarly, these seven clusters consistently found a high similarity match in networks with different target link densities (Supplementary Information Figure 1b, averaged Weighted Jaccard Similarity = 0.81). The small clusters (<1% of the population) were more idiosyncratic to the specific network thresholding parameters. Hereafter, we refer to these seven stable clusters of people as "Creative Species" to represent different themes in creative habits that tend to be found together.

While there were not strong correlations between the ordinal responses of individual question pairs, the Creative Species clusters suggest that there were significant multi-dimensional combinations of creative habits that co-occur (Figures 1b, and 2). Table 2 summarizes the creative habit sets that were the most common or most over-represented in each Creative Species, and Box 1 provides narrative descriptions. It is important to note that the most representative creative habits of each Creative Species are not mutually exclusive. For example, there are three different Creative Species where "Novelty Seeker" was over-represented (at least 50% more common in the cluster than in the overall population) (Table 2). The other co-associated top habits, however, differed between groups to create three different "flavors" of novelty-seeking Creative Species. 86% of all creative habits were over-represented in at least one Creative Species cluster, and 52% were over-represented in at least two. Only six creative habits ("Quite/Silence", "Generalist", "Tidy Workspace", "Pragmatist", and "Perfectionist") were not over-represented in any Creative Species cluster. These habits were not unusually common or rare overall (Figure 1a, range 23-38%), rather they were more randomly associated with other creative habits spanning different Creative Species.

To compliment the patterns observed in the network of people linked if they share similar sets of creative habits (Figure 2), we generated a network of creative habits, linked if they co-occur in similar sets of people (Figure 3). This creative habit network highlights the strongest co-occurrence trends across all creative habit pairs, and it revealed three clusters of creative habits that were more likely to be found in the same sets of people than expected by chance (Figure 3, colored clusters). The largest of these habit clusters was characterized Risk friendly, Novelty Seeker, Multitasker, Stifled by Constraints, Reframer, and Internally Motivated as the most central, or archetypal, in the group (Figure 3, blue nodes). In the second largest habit cluster, the most archetypal habits were Monotasker, Solo Creator, Private Workspaces, Risk Averse, and Self-Critical (Figure 3, orange nodes). The smallest habit cluster was characterized by Rational,

Hate Distractions, Stimulated by Constraints, Externally Motivated, and Outwardly Inspired (Figure 3, yellow nodes).

Of the subset of respondents who provided data about their profession, 83% were in Arts & Design, 52% were in Science & Engineering, and 35% were in Business & Entrepreneurship (note that the sum is greater than 100 because an individual could have more than one discipline). Each discipline was well represented in all seven Creative Species, and vice versa (see http://bit.ly/3c5aDO1 for an interactive visualization to select each discipline). The strongest pattern of over- and under-representation of a discipline in some Creative Species was by people in Business & Entrepreneurship, who were almost 50% more common in Creative Species 3 than expected by chance (Figure 4). This Creative Species was characterized by people who prefer to do highly collaborative work and are creatively stimulated by risk. People in this Creative Species also tended to consider physical movement to be an important part of their creative process, and they were more likely to manage obstacles by reframing the problem than by tenaciously sticking to 'Plan A' (Table 1). Businesspeople were 20%-33% less common in Creative Species 1, 6, and 7 than expected by chance. These Creative Species were characterized by risk-aversion, routine-seeking, and the tendency to be creatively stifled by constraints (Figure 4, Table 1). Artists and Scientists were more evenly distributed across all Creative Species. While they showed some over- or under- representation in certain Creative Species (Figure 4b); the absolute magnitude of the differences was never more than 10% of the chance expectation (Figure 4a).

Of the 70% of the respondents who provided data about their gender, 59.8%, 39.7%, and 0.5% self-identified as male, female, and non-binary/non-conforming, respectively. Female respondents were over-represented in Creative Species' 1 and 6, while Male respondents were over-represented in Creative Species 3 and 5 relative to the permuted chance expecation (Figure 5b); however, the magnitude of the differences was generally less than 10% (Figure 5a). Non-binary/non-Conforming respondents showed larger absolute variation in their prevalence across the seven Creative Species relative to expected (5a), but the low permuted z-scores suggest these trends were an artifact of low sample size (Figure 5b).

**Discussion:**

We developed a survey that allowed characterization of 42 "creative habits" along 21 dimensions of preferences and tendencies that people have in their creative process. Each dimension identifies two opposing creative habits at each endpoint of the scale (e.g., "Monotasker" vs. "Multitasker"). Similar to research on cognitive styles, neither creative habit endpoint was 'good' or 'bad' relative to creativity[10]. Instead, most habit pairs represent tradeoffs in preferences or tendencies which may be advantageous for some contexts but disadvantageous in others. Similarly, different habits may lend themselves to processes that operate in different stages of the creative process[11–13] For example, Multitasking may enhance creativity by facilitating divergent thinking, reducing inhibition, allowing for cross fertilization of solutions, or allowing for subconscious incubation on one project while working on another[8,14,15]. Similarly, working on multiple projects at once may foster creativity by building "networks of enterprise."[6,16] At the same time Monotasking may help the in-depth thinking that allows remote associates and other good ideas to be found; and Monotaskers benefit from being less easily distracted, more productive, and less error-prone than Multitaskers[17]. Thus, Multitasking may be more helpful in the early ideation phase of a project, while Monotasking may be more advantageous in the execution phase.

It is important to note that the creative habits on any given dimension are not mutually exclusive. For example, a person could prefer multitasking at certain times (some projects, or certain phases of the creative process) and monotasking under other conditions. However, the questions were designed to capture people's strongest general preferences. Creative habits of either endpoint were only assigned if the response was stronger than the median response score of the population. If a respondent selected the mid-point of the response scale, it was assumed that either they had no strong preference/tendency, or it was highly variable depending on the circumstance. 90% of the respondents expressed strong enough general preferences to be assigned at least three creative habits, and most respondents had strong enough preferences to be assigned at least 14 creative habits.

While each question was statistically independent (the average absolute rank correlation in ordinal scores across all question pairs was 0.03, and the maximum 0.38), there were clear higher dimensional patterns in the combinations of creative habits that tend to be found together in the same sets of people. We identified seven "Creative Species", or clusters of people with higher similarity in their creative habits than expected by chance, which were stable to both random sub-sampling and variation in link thresholding (Figure 1, SI Figure 1). The defining creative habits in each Creative Species were not unique. 36 of the 42 creative habits were over-represented in at least two clusters, and four creative habits (Stifled By Constraints, Internally Motivated, Solo Creator, and Novelty Seeker) were over-represented in three clusters. Differences in other co-associated creative habits characterized the diversity of these Creative Species (Table 2, Box 1).

Patterns in the overlap among Creative Species clusters in their top habits are reflected in the spatial arrangement of the most archetypal people of each Creative Species cluster (Figure 2). The clusters self-sort horizontally along a broad gradient from more "deliberate" Creative Species on the left to those with more "open" Creative Species on the right. The over-represented creative habits in the left-most clusters (Figure 2, Creative Species 1, 6, and 7) included Solo Creator, Private Workspaces, Dislike Distractions, Risk-Averse, Routine-Seeker, Quiet/Silent Work Environment, Monotasker, Tenacious, and Make It Happen. In contrast, the most over-represented creative habits on the right (Figure 2, Creative Species 2, 3, and 5) included Novelty Seeker, Risk-Friendly, Multitasker, Reframer, and Let It Happen. There was also a weaker vertical gradient broadly defined by "outward/rational" vs "inward/intuitive" creators. The lower region of the inverted "U-shape" (Figure 2. Creative Species 1, and 5) was more likely to include the creative habits Externally Motivated, Outwardly Inspired, Stimulated by Constraints, and Rational. The upper part of the inverted U in Figure 2 was less defined but was more likely to include people with the opposing creative habits, such as, Internally Motivated, Internally Inspired, Stifled by Constraints, and Intuitive.

These gradients across the seven Creative Species clusters (Figure 2) are reflected in the second, "creative habit network" which links creative habits if they are more likely to co-occur in the same sets of people (Figure 3). The two largest creative habit clusters in Figure 3 highlight "open" (Creative Species 1: Risk Friendly, Novelty Seeker, Multitasker, Let it Happen) vs "deliberate" (Creative Species 2: 'Solo Creator, Monotasker, Private Workspace, Routine Seeker) sets of habits respectively.  The third and smaller creative habit cluster reflects the weaker, vertical dimension of Creative Species described above where the most archetypal habits include Rational, Externally Motivated, Outwardly Inspired, and Stimulated by Constraints. Note that Dislike Distractions and Quiet/Silent Workspace are in this cluster, but they are strongly cross-linked with numerous habits in the 'deliberate' habit theme. This habit network also shows which creative habits do not have strong or consistent co-associations with others. For example, "Pragmatist" and "Perfectionist" are peripheral, not because they are rare, but because they only weakly co-occur (thin lines) with other creative habits, and those habits fall in different groups. In other words, having a more pragmatic vs perfectionist approach to one's creative work is not a good predictor of their other creative habits or their Creative Species.

Each broad discipline (Arts, Business, Science) included people that spanned all Creative Species.  Artists and Scientists were more evenly distributed across Creative Species than Businesspeople (Figure 4a). This observed broad, and overlapping, creative diversity of Artists and Scientists is interesting given the longstanding debate about Art vs Science representing two distinct "cultures"[8,18]. While Artists (including designers, graphic artists, performers, etc.) and Scientists (including researchers, engineers, technologists, etc.) were more (or less) common in some Creative Species than expected by chance (Figure 4b: z-scores >3), the absolute magnitude of the differences were never more than 10% of the expected value from 1000 permutations of the data (Figure 4a).  People in Business (including entrepreneurs, finance, etc.) had the strongest patterns of Creative Species preference.  They were 48% more common in Creative Species 3 than expected by chance (Figure 4a). This group was characterized by "open" and "outward/rational" creative habits (e.g, Collaborator, Risk Friendly, Multitasker, Reframer, Kinetic; Table 2, Box 1).  Businesspeople were also strongly under-represented in Creative Species 1, 6, and 7 (Figure 4), all of which are on the left side the Creative Species archetype network (Figure 1) and are characterized by more "deliberate" types of creative habits (e.g., Solo Creator, Private Workspaces, Hate Distractions, Risk Averse, Routine Seeker, Silent Work Environment, Monotasker, and Make it Happen; Table 2, Box 1). It is important to note that even if some disciplines were more commonly found in some Creative Species than expected by chance, all broad disciplines were found in all Creative Species and vice versa – and knowing someone's discipline alone is not necessarily a good predictor of their creative preferences or tendencies.

Each Creative Species included people from all genders and vice versa. Even the small set of respondents which self-described as "non-binary/non-conforming" were present in all seven Creative Species clusters.  There was a trend for female respondents to be over-represented in two Creative Species (1 and 6) characterized by more "deliberate" creative habits, and for male respondents to be over-represented in Creative Species 3 and 4 characterized by more "open" creative habits (Figure 5b, Table 2, Box 1).  However, the absolute magnitude of these differences relative to the permuted mean was small (Figure 5a); and knowing someone's gender is not a good single predictor of their creative preferences.  Non-Binary/Non-Conforming respondents showed a larger variation relative to the permuted expectation, but the small z-scores suggest this was an artifact of small sample size (Figure 5b).

Together these results suggest that independent of gender or discipline, people vary widely in the habits, behaviors, and contexts in which they feel most creative. While the present study is, to our knowledge, the largest empirical study of creative habits conducted, we underscore here some caveats in the scope of inference of this work.  Our sample was limited to <10,000 people who participated in a survey on a public-facing website. The survey focuses on 42 creative habits along 21 independent dimensions.  A broader understanding of the variety of ways in which people create would benefit from more explicitly recruiting and quantifying broader racial, ethnic, socio-economic, and demographic diversity in survey participants, and from expanding the survey to include other creative habits and preference dimensions deemed important in creative work.  Expanding our collective understanding of Creative Diversity is critical for helping us all cultivate our creative strengths, be mindful of their trade-offs, and successfully collaborate with others who are different from ourselves.

**Methods:**

*Survey Development Phase 1. Crowd-sourced testing*: The first version of the survey was 270 questions (Supplementary Information, Table 1) which spanned a broad range of question types that included 'pick one', 'check multiple', ordinal (1-5 scale), and open-ended (free text). The survey questions explicitly covered all phases of the creative process: problem definition, ideation, evaluation, and implementation; as well as 'creative habit' themes commonly mentioned in the popular press. It also included questions that were related to personality tests, but with an emphasis on creativity. We then tested this survey on 1,000 people using the micro-payment crowd-sourcing platform CrowdFlower (now https://appen.com). Participants were also offered bonus payments to participate in a follow up survey to provide feedback on the questions and their experience.

*Survey Development Phase 2. Curated testing and refinement*: We used the results and feedback from the first phase to reduce redundancy, streamline questions, and refine question phrasing. We also removed questions from Phase 1 that were focused more on personality type than explicitly on creativity. This process narrowed the survey down to 91 questions (Supplementary Information Table 2), also of mixed type. This survey was administered to a community of 500 people curated by the Red Bull's "High Performance Program" group which trains sponsored athletes and other elite performers. The curated group of participants were actively chosen to represent a wide range of disciplines, but with a common theme that they had demonstrated publicly recognized achievement in their professional endeavors. Examples of the range of survey participants include: Rodney Mullen (founder modern street skateboarding), Powtawche Valerino (NASA rocket scientist), Nelson Dellis (four-time USA Memory Champion), Kristin Allen (Hall of Fame American acrobatic gymnast), Benjamin Dupont (Creative Director for Cirque du Soleil), Sonny John Moore (known as the DJ, musician, and producer Skrillex), Tyler Florence (celebrity chef), and Stephen Kearin (improvisation performer). We do not present these results in this survey here because a) the sample size was too small to identify stable 'Creative Species', and b) the survey was different enough from the final survey to preclude combining the results. The results of this survey were used to motivate public participation in the final, shorter survey (*Survey Development Phase 3*).

*Survey Development Phase 3. Final survey with broad public participation*: The results of the second phase were used to further streamline and simplify the survey by a) removing/merging 'redundant' questions which had strong co-variation in their responses and b) converting as many questions as possible to a simple 5-point Likert scale where each question addressed an independent dimension of creative habits. The final survey (Table 1) was 21 questions, where the two ends of the response scale were mapped to one of two possible creative habits which represented endpoints of a potential tradeoff along a given preference dimension (e.g., 'Mono-tasker' vs 'Multi-tasker' for the question "On average, how many creative projects do you juggle simultaneously?"). To engage broader public participation, the results from the curated community of 500 'high performers' were displayed on a public website to set the context for the survey. Each of the 500 people had an accessible profile with their results, their short biography, links to their websites, social media feeds, and other inspirational media that they uploaded in the Phase 2 survey. Visitors to the website could take the short 'Creative Species Survey' and their responses were used to match them to the top 3 'high performers' with the most similar creative habits. Since the 'high performer' results were based on the longer 91-question survey, we mapped the responses for each question in the longer survey to those of the short survey. These mapped results were then used to match new respondents with their most similar 'high performers.' This approach helped incentivize participation and survey completion. Of the 10,637 participants that logged on to the survey, 9,647 (91%) completed at least 90% of the questions. This study reports the result of this final survey with broadest public participation.

*Data Analysis:*

   a) *Assigning Creative Habits* – Each question was designed to assess if a respondent has strong preference or tendency for one of two "opposing" creative habits. 20 of the 21 questions were on a 1-5 ordinal Likert scale, with the exception being a question about one's most creatively productive time of day. The latter was a 'check all that apply' question with options to choose different times of day, but the final responses were mapped to one of two endpoints ('Early Bird' vs 'Night Owl'). For each question, we assigned the respondent one of the two creative habits if they had a strong preference for one or the other endpoint of the scale. We interpreted the middle value (3) to be either 'no strong preference' or 'it depends'. For most questions, the scores were normally distributed around the middle value (3). In these cases, the person was assigned one of the creative habits if they selected either extreme (1 or 5). In cases where the responses were skewed and the median score was greater than 3, they were assigned the creative habit on the lower end of the scale if they selected 1 or 2. If the responses were skewed in the other direction, they were assigned the creative habit on the right end of the scale if they selected 4 or 5. Thus people were assigned a creative habit if their preference for that endpoint was strong relative to the rest of the population.

   b) *Identifying "Creative Species"* – To identify broad themes in creative habits, we clustered people into groups that tend to share similar combinations of habits using a network analysis approach previously used to cluster documents with similar keyword sets. Details of this approach can be found in Halpern et al.[19] and Williams et al.[1], and the python scripts are open source (https://github.com/foodwebster/Tag2Network). In our case, we built a network of people linked based on similarity in their creative habit sets. We restricted the analysis to 9,633 participants who had at least three creative habits, which was 99.85% of the those that answered at least 90% of the survey questions. Clusters of people with similar creative habits were found using the Louvain community detection method[20]. This algorithm produces a hierarchical clustering of nodes into groups which are more densely connected with each other than expected by chance. We used a directed version of the Louvain method[21] because linking entities to their top *n* most similar neighbors has a direction. In other words, you may be my most similar neighbor, but I may not be yours.

We used each cluster to define a different 'Creative Species', a group of people with a set of distinctively similar creative habits. To characterize the most distinctive creative habits of each 'Creative Species' we summarized the distribution of creative habits across people in the group. The 'top creative habits' for each Creative Species were defined as those that were both common (present in at least 60% of the people of group) and over-represented (at least 1.5 times more frequent in the group than in the total population of respondents) (Table 2).

To visualize patterns in top creative habits across the Creative Species, we subset network to include the respondents who were the most 'archetypal' of each cluster. Each person was assigned a 'cluster archetype' score, which measures the degree to which it is a canonical representative of its cluster[1]. The cluster archetype score is normalized across each cluster, so we selected all network nodes with a score greater than zero to represent all people that are more archetypal than the average person in the cluster. This subset of 3,154 people was then visualized as the "Creative Species Archetypes Network" (Figure 1), with an interactive version accessible online at http://bit.ly/3c5aDO1.

To help interpret the Creative Species clusters – which are groups of people that tend to share similar sets of creative habits – we used the same similarity network generation method to create a network of creative habits, linked if they co-occur in similar sets of people. In this "Creative Habits Network", the nodes are habits (not people) and they are clustered into groups of habits that tend to be found together in the same people (Figure 2). An interactive version is accessible online at http://bit.ly/3wHAIdH.

For both networks, to visualize the relationships between clusters in space, we use t-Distributed Stochastic Neighbor Embedding (Python scikit-learn version 0.24.1, sklearn.manifold.TSNE) which attempts to preserve in 2-dimensions at multiple scales the relative distance between entities in n-dimensional space. Here we used the shortest path length between each node pair as the distance metric. This approach reveals patterns at multiple scales, where nodes which are more densely inter-linked are closer to one another in space, and clusters which have more cross-links between them are more likely to be adjacent to one another.

    c) *Sensitivity Analysis* – We evaluated the stability of the Creative Species clusters in two ways. First, comparing the clusters to those generated from 100 random sub-samples of the population. For each trial, 50 percent of the population was randomly sampled with replacement; a network was generated; and clusters were identified as described above. For each sampled network, each cluster was compared to every cluster in the original full network by computing the Weighted Jaccard Similarity of creative habits between each cluster pair. For each cluster in the original network, we identified its most similar cluster in each sampled network (i.e. the cluster with the highest overlap in its creative habit frequency distribution) and summarized the average and standard deviation of maximum similarity across all 100 trials (Supplementary Information, Figure 1a).

    d) Our second approach was to evaluate how sensitive the largest clusters (those with at least one percent of the population) are to variation in the thresholding parameter for network generation. Since a similarity network is built by thresholding a similarity matrix to create a sparse network, we varied the final target link density from 6 to 12 links per node to generate seven networks from the full dataset and compared the clustering results. The number of large clusters with at least one percent of the population decreased with increasing link density from 11 to 6 clusters – and plateaued at 7 clusters for networks with 9, 10, and 11 links per node. We used these results to choose 10 links per node as our optimal thresholding parameter and then compared all the large clusters in this focal network to those in the rest. Each cluster in the focal network consistently found a high similarity match to a cluster in the other networks with

differing link density (average Weighted Jaccard Similarity to each best match was 0.80) (Supplementary Information, Figure 1b).

e) *Analyzing Disciplines and Gender* — At the end of the survey, we asked respondents to provide information about their gender and discipline. 70% of the respondents provided information about their gender, and these responses were bucketed into 3 categories: Male, Female, and Non-Binary/Non-Conforming. The discipline question asked "Which of the following might be used to describe your professional or creative work?" with the option to check as many disciplines as they want from a list or check 'other' and add a free text response. 66% of the respondents answered this question. 91% of these responses (or 60% of total responses) could be mapped to the three most common broad disciplines: "Arts & Design" (visual arts, performing arts, music, graphic design, etc), "Business & Entrepreneurship" (business, entrepreneurship, social enterprise, finance, etc.) and "Science & Engineering" (science, research, technology, engineering, medicine, biology, etc.). These broad disciplines (hereafter be referred to as "Arts" or "Artists", "Business" or "Businesspeople", and "Sciences" or "Scientists", respectively) were not mutually exclusive (i.e., a respondent could be assigned more than one).

For both discipline and gender, we conducted a permutation test using the subset of the population for which we had data to explore whether a given broad disciplines or gender was surprisingly over- (or under-) represented in some Creative Species over others. For each discipline (or gender), we calculated the observed fraction of respondents in each cluster. We then randomly shuffled the discipline (or gender) labels and calculated their distribution of across Creative Species in the permuted data. We repeated this 1000 times to compute the mean (and standard deviation) fraction of each discipline (or gender) in each Creative Species expected by chance. A z-score for the observed fraction of each discipline (or gender) in each cluster was calculated as (($observed - permuted\ mean$) / $permuted\ standard\ deviation$). The magnitude of the difference was measured as ($observed / permuted\ mean$)*$100$ (Figures 4 and 5).

**References:**

1. Williams, R., Runco, M. A. & Berlow, E. Mapping the Themes, Impact, and Cohesion of Creativity Research over the Last 25 Years. *Creativity Research Journal* **28**, (2016).
2. Simonton, D. K. Creativity as Blind Variation and Selective Retention: Is the Creative Process Darwinian? *Psychological Inquiry* **10**, 309–328 (1999).
3. Gruber, H. E. The Life Space of a Scientist: The Visionary Function and Other Aspects of Jean Piaget's Thinking. *Creativity Research Journal* **9**, (1996).
4. Tweney, R. D. Presymbolic Processes in Scientific Creativity. *Creativity Research Journal* **9**, (1996).
5. Wallace, D. & Gruber, H. E. *Creative People at Work: Twelve cognitive case studies*. (Oxford University Press, 1989).
6. Root-Bernstein, R. S., Bernstein, M. & Garnier, H. Correlations Between Avocations, Scientific Style, Work Habits, and Professional Impact of Scientists. *Creativity Research Journal* **8**, (1995).
7. Ludwig, A. M. *The Price of Greatness*. (Guilford Press, 1995).
8. Root-Bernstein, R. Art/Science. in *Encyclopedia of Creativity* (Elsevier, 2020). doi:10.1016/B978-0-12-809324-5.23719-X.
9. Plucker, J. A. Beware of Simple Conclusions: The Case for Content Generality of Creativity. *Creativity Research Journal* **11**, (1998).
10. Kirton, M. *Adaptors and Innovators: Styles of creativity and problem solving*. (Routledge, 1989).
11. Sadler-Smith, E. Wallas' Four-Stage Model of the Creative Process: More Than Meets the Eye? *Creativity Research Journal* **27**, (2015).
12. Allen, A. P. & Thomas, K. E. A Dual Process Account of Creative Thinking. *Creativity Research Journal* **23**, (2011).
13. Runco, M. A. & Basadur, M. Assessing Ideational and Evaluative Skills and Creative Styles and Attitudes. *Creativity and Innovation Management* **2**, (1993).
14. Loh, K. K. & Lim, S. W. H. Positive associations between media multitasking and creativity. *Computers in Human Behavior Reports* **1**, (2020).
15. Collins, M. J. D. A Distracted Muse: The Positive Effect of Dual-Task Distraction on Creative Potential. *Creativity Research Journal* **32**, (2020).
16. Gruber, H. E. Networks of enterprise in creative scientific work. in *Psychology of Science: Contributions to Metascience* (ed. Gholson, B.) (Cambridge University Press, 1989).
17. Ophir, E., Nass, C. & Wagner, A. D. Cognitive control in media multitaskers. *Proceedings of the National Academy of Sciences* **106**, (2009).
18. Feist, G. J. Synthetic and analytic thought: Similarities and differences among art and science students. *Creativity Research Journal* **4**, (1991).
19. Halpern, B. S. *et al.* Ecological Synthesis and Its Role in Advancing Knowledge. *BioScience* (2020) doi:10.1093/biosci/biaa105.
20. Blondel, V. D., Guillaume, J.-L., Lambiotte, R. & Lefebvre, E. Fast unfolding of communities in large networks. *Journal of Statistical Mechanics: Theory and Experiment* **2008**, P10008 (2008).
21. Dugué, N. & Perez, A. *Directed Louvain : maximizing modularity in directed networks*. https://hal.archives-ouvertes.fr/hal-01231784 (2015).

**Acknowledgements:**

The ideas in this work benefited greatly from many current and former collaborators. In particular we would like to thank (in alphabetical order): Kristen Allen, Jessica Green, Sean Gourley, Amy Heineike, Oliver Holzmann, Kristin Kuzemko, Sundev Lohr, Christopher Lortie, Per Lundstam, Rodney Mullen, Ben Potvin, Norman Seef, Skrillex, Aditya Vishwakarma. This work was funded in part by Red Bull North America Inc. and Logitech Inc. The funders had no role in study design, data collection and analysis, decision to publish, or preparation of the manuscript.

**Author Contributions**

ELB led the analysis and writing of this manuscript. AW and AC conceived of and managed the original Hacking Creativity project. ELB and RJW led the data analysis. DG led design and KD led technical execution for the Hacking Creativity project. LS and MR led the survey design, development, and refinement. SC, SJ, and TY developed the botanical being metaphors and designed the public facing website related this work. All coauthors contributed significantly to the ideas, study design, and interpretation of results.

**Competing financial interests:** The authors declare no competing financial interests.

**Box 1. Seven "Creative Species"**

We identified seven "Creative Species," or clusters of people with higher similarity in their creative habits than expected by chance (Figure 2). Table 2 summarizes the most common and over-represented creative habits in each cluster. Table 1 provides the most literal interpretation of how each creative habit maps to a specific survey question. Here we provide more metaphorical descriptions of each as a hypothetical botanical species. We chose this botanical metaphor because each plant species on earth is a unique creative solution to the universal problem of life – how to survive, grow, and reproduce. Every unique combination of plant forms and strategies for solving this problem has its own advantages and trade-offs, and each has its own set of environments and habitats in which it thrives (or does not thrive). This metaphor also acknowledges that while individual plants can be grouped into species and tend to thrive in certain types of places, there still exists large variability among individuals within a species in their phenotype and habitat preferences. For a more immersive exploration of these metaphorical Creative Species, see www.thecreativelandscape.com.

**Creative Species 1:** *Mono routinus*
Top 5 creative habits: Risk Averse, Routine-Seeker, Dislike Distractions, Monotasker, Make it Happen.
*Mono Routinus* typically grows its best blooms while alone, preferring enclosed environments where it can find predictable conditions in which to spread its roots. Similar to *Solo noctus*, this creative species does not take chances with its precious resources, and only begins growing a bloom when all conditions are perfect. Those who come across a *Mono routinus* in the wild should be respectful of its privacy and careful not to disturb its process. Although it does not produce flowers at the speed of some other species, if given the right conditions, *Mono routinus* will reliably produce perfect blooms throughout its remarkably productive life

**Creative Species 2:** *Yolo chaotis*
Top 5 creative habits: Novelty-Seeker, Risk Friendly, Stifled By Constraints, Internally Motivated, Multitasker.
*Yolo chaotis* is best known for its many, varied, and sometimes wild thickets of bright flowers. A novelty-seeker like Novo Gregarious, this creative species thrives in mountain streams, where the current can carry it to new and changing environments. Unlike *Socialis adventurous*, which draws nutrients from its environment, *Yolo chaotis* haotis has all it needs to thrive contained within its bulb, and will spread its energy across many new flowers at once, even if conditions aren't perfect. Since *Yolo chaotis* is inwardly-energized, in the right conditions the careful observer may even see a subtle glow emanating from deep inside this special creative species.

**Creative Species 3:** *Socialis adventurous*
Top 5 habits: Collaborator, Risk Friendly, Multitasker, Reframer.
*Socialis adventurous* is characterized by its many blooms that take flight to ride air currents in search of new landscapes. A highly-social creative species, *Socialis adventurous* is found in habitats dense with others, its roots flowing out to collect a wide range of nutrients, the more the better. Like *Yolo chaotis*, *Socialis adventurous* liberally spreads its energy across many blooms at once, knowing that even though not all of them will flourish, those blooms that do, will be strong and vivid. Those who would like to see the best of this creative species will need to rise early, because many *Socialis adventurous* prefer to show their brightest colors in the morning.

**Creative Species 4:** *Focus mononovus*
Top 5 creative habits: Novelty-Seeker, Monotasker, Stifled By Constraints, Silence, Dislike Distractions.
*Focus mononovus* is a detailed and dignified creative species that, like *Novo gregarious*, travels streams and rivers in search of novel, unconstrained environments. Single-minded and precise, *Focus mononovus* requires calm and quiet to focus its energy into growing a colorful, perfectly-formed bloom. Although many creative species can thrive in indoor environments, *Focus mononovus* grows best in the wild, unconstrained and free. Those who find a *Focus mononovus* in the wild should be careful not to distract it, otherwise it may have to start its flowering process over again.

**Creative Species 5:** *Novo gregarious*
Top 5 creative habits: Novelty-Seeker, Externally Motivated, Outwardly Inspired, Collaborator, Noise/Music.
*Novo gregarious* is a highly-social creative species that thrives in noisy, bustling environments, its roots floating downstream to absorb nutrients, and the abundant energy emanated by the species around it. Community is *Novo gregarious*' primary motivating force, so it often does not bloom until encouraged to do so by others around it. A vibrant and joyful species, *Novo gregarious* has the unique ability to bring together species that otherwise would bloom alone. Creative species spotters who would like to see a *Novo gregarious* bloom, should bring a friend along, play music, and can encourage its flowering with a lively conversation.

**Creative Species 6:** *Sui inspira*
Top 5 creative habits: Routine-Seeker, Private Workspace, Internally Motivated, Solo Creator, Inwardly Inspired.
*Sui inspira* is found in quiet, enclosed habitats, protected from passersby and far from other flora and fauna. This unique and magical creative species requires a predictable environment and has a deeply-rooted bulb that contains all the nutrients it needs to grow. Because it is so self-sufficient, *Sui inspira* is happy in isolation, diligently growing its intricate, gently-glowing bloom. Those who come across a Sui Inspira in the wild should be quiet, respectful, and let it bloom in peace.

**Creative Species 7:** *Solo noctus*
Top 5 creative habits: Stifled By Constraints, Internally Motivated, Solo Creator, Night Owl, Risk Averse.
*Solo noctus* can be found in open terrain blooming under a wide-open night sky and will not thrive in constrained environments. Like *Sui inspira*, this creative species grows from a bulb that contains everything it needs to succeed. *Solo noctus* is tenacious and dedicated, waiting until conditions are perfect to put all its energy into growing a single, densely-petaled flower. A *Solo noctus* does poorly in captivity, so those who want to see its fabled flower in person should be prepared to stay up late, because this creative species only blooms at night.

FIGURES

Figure 1. The distribution of a) creative habits and b) Creative Species clusters across 9,633 people. Note that none the of the creative habits were so ubiquitous as to be in the majority of respondents, and all were present in at least 10% of the population The colored bars in (b) are the top seven largest and most stable 'Creative Species', which are defined as clusters of people that tend to share similar combinations of creative habits. The defining creative habits of each cluster can be found in Table 2. The grey bar includes 23 small and unstable clusters which together comprise less than 1.7% of the total population of respondents.

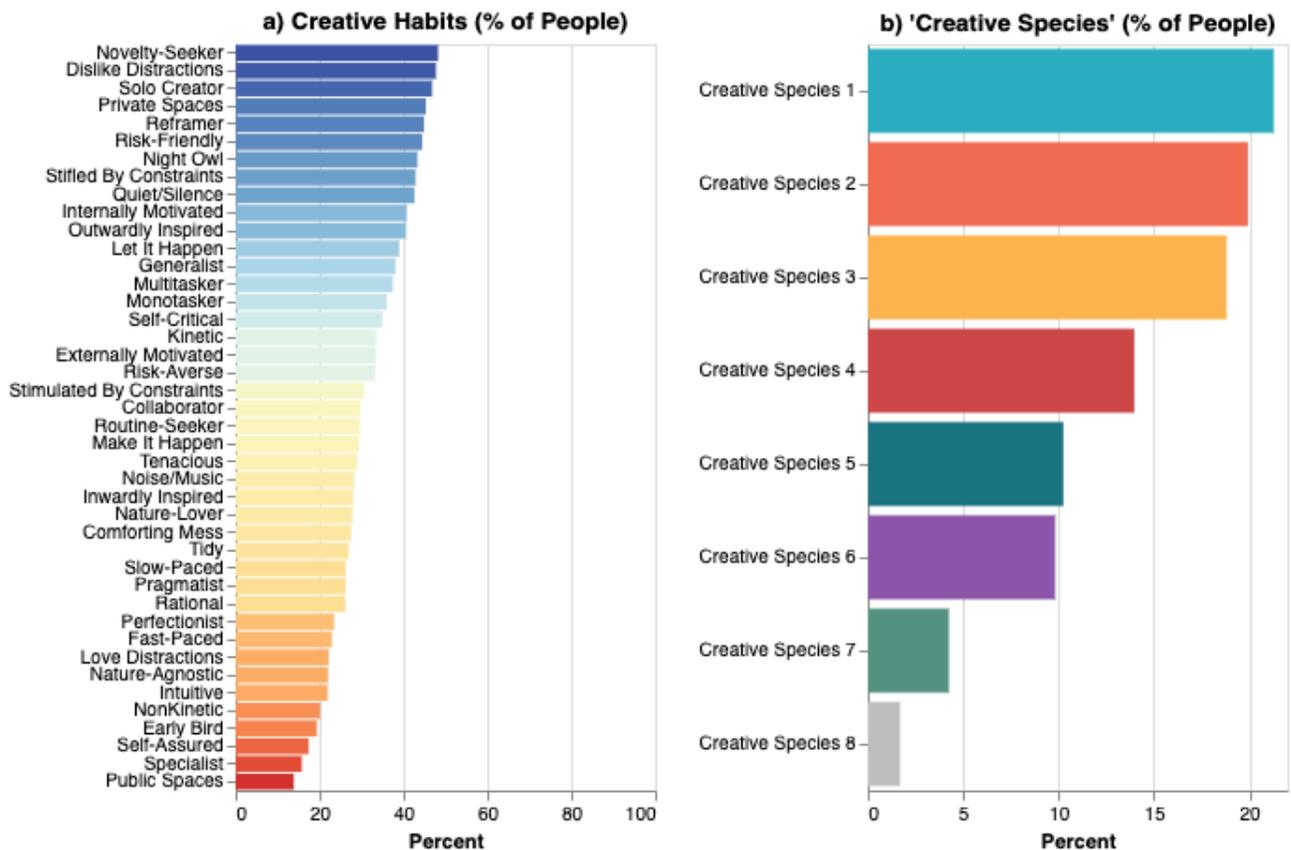

Figure 2. a) Network of the most archetypal people in each Creative Species cluster. Colors are the same as in Figure 1b, and the defining creative habits of each cluster can be found in Table 2 and Box 1. Each node is a respondent, and they are linked if they share similar sets of creative habits. The colored clusters are emergent "Creative Species" defined as groups of people with similar combinations of creative habits. Larger nodes have a higher cluster archetype score. The t-SNE spatial layout places nodes and clusters which are more interlinked closer to one another in space. b) Examples of the left-right gradient from more 'deliberate' to more 'open' creative habits. Highlighted nodes show people who are more creative when monotasking vs multitasking, or when they have a routine vs when their surroundings are novel and varied. Other trends can be explored in the interactive version of this network at http://bit.ly/3c5aDO1.

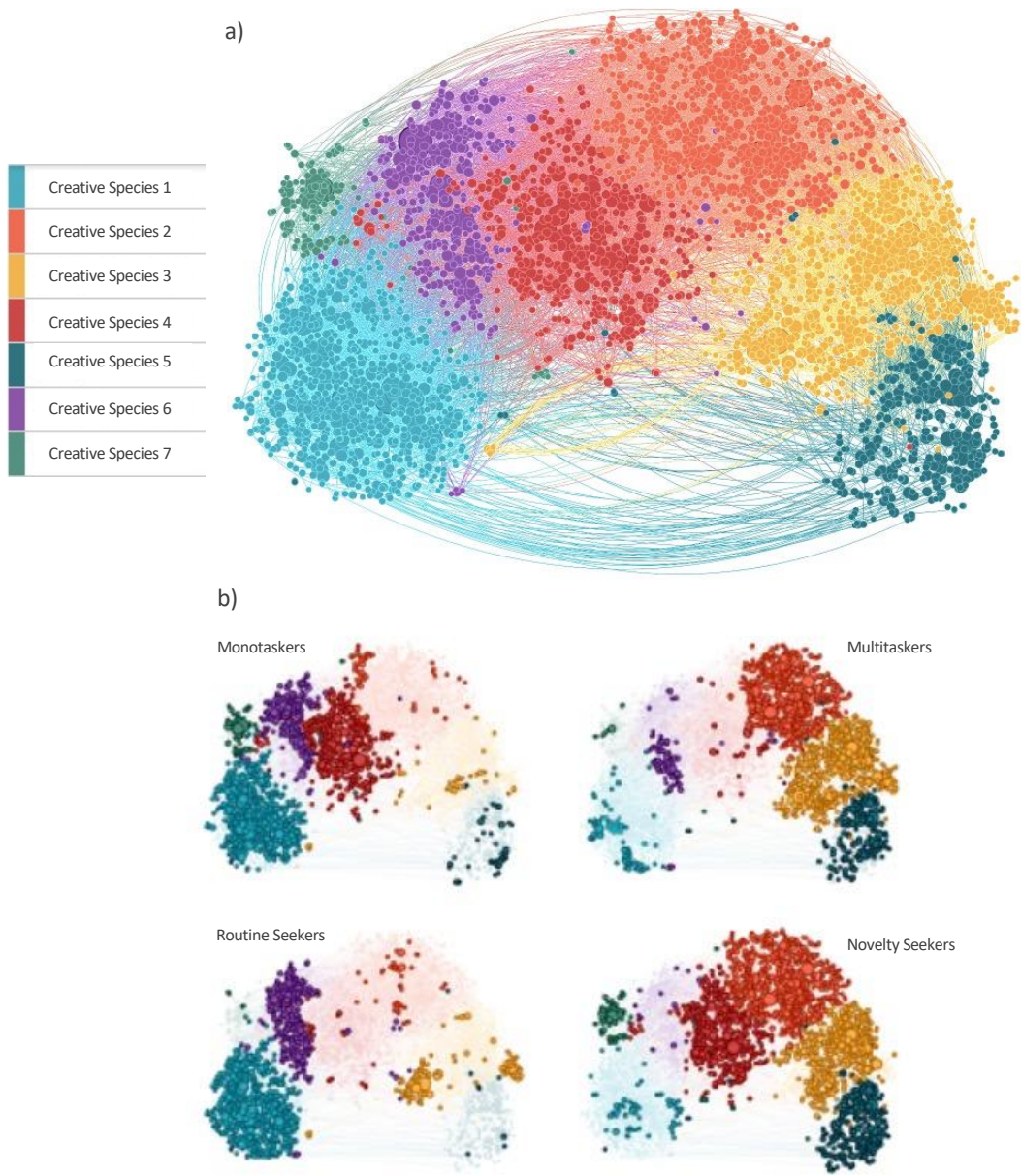

Figure 3. Network of Creative Habits linked if they co-occur in similar sets of people. The colored clusters - or 'Habit Themes' - are groups of creative habits that tend to co-occur with one another more than expected by chance. Nodes are sized by their cluster archetype score (larger ones have many co-associations with others in the same cluster). Each Habit Theme is labeled by the top three most archetypal, or central, habits of the cluster. Thicker links indicate stronger co-association between a given pair of creative habits. The layout of nodes and clusters was determined with t-SNE so that nodes and clusters which are more interlinked are closer to one another in space. An interactive version of this network can be accessed at http://bit.ly/3wHAIdH.

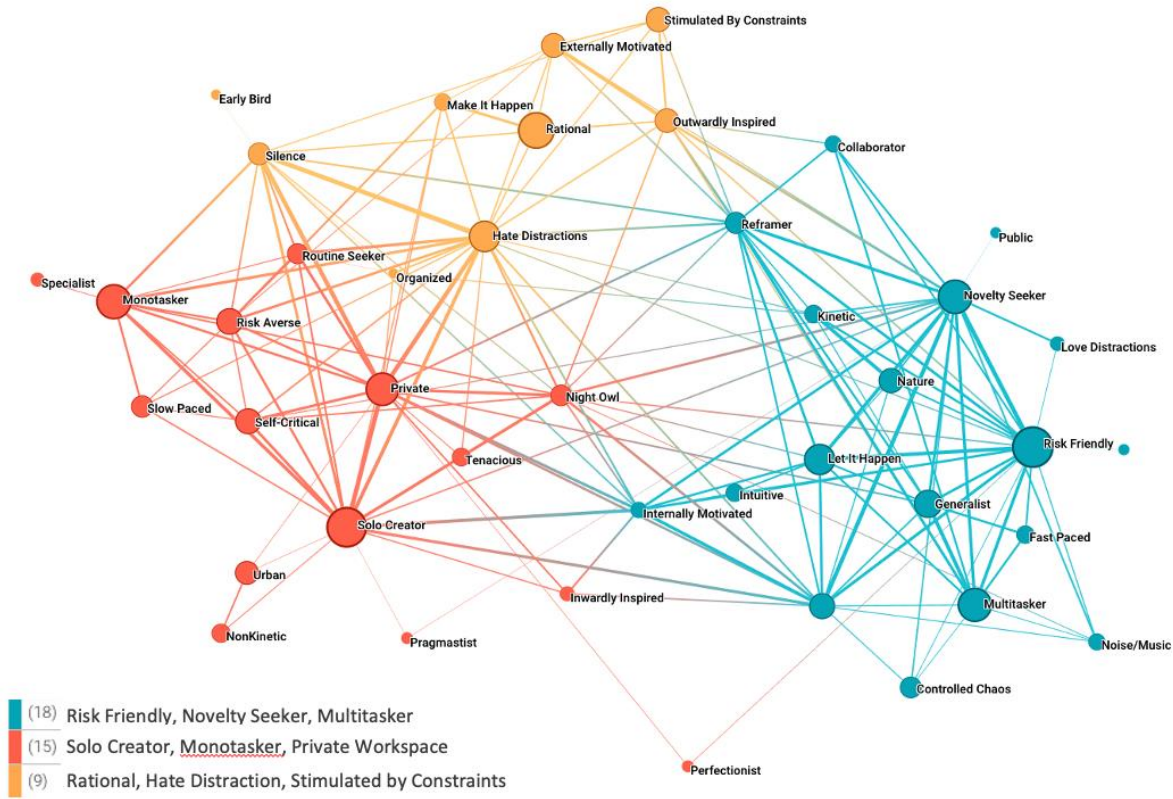

Figure 4. Over- vs under-representation of Creative Species by broad discipline ("Arts-Design", "Business-Entrepreneur", "Science-Engineering"). Colors are the same as in Figure 1b, and the defining creative habits of each cluster can be found in Table 2. a) The percent difference between the observed and permuted average fraction of people in each broad discipline per Creative Species cluster. b) The z-score (number of standard deviations from the permuted average) for the observed fraction of people in each broad discipline per Creative Species cluster. Both the percent difference and z-scores for each cluster and discipline were calculated relative the mean fraction of people in that discipline in 1000 random permutations of the discipline labels. We considered a discipline to be significantly over- (or under-) represented in a cluster if the z-score was greater than 3 (or less than -3) standard deviations of the mean frequency in 1000 equal sized random samples.

a)

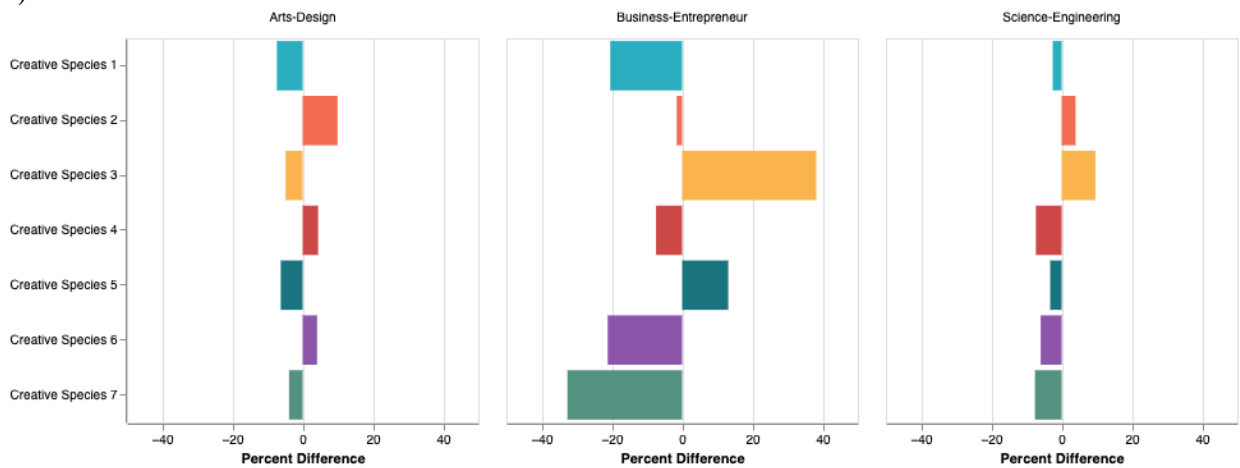

b)

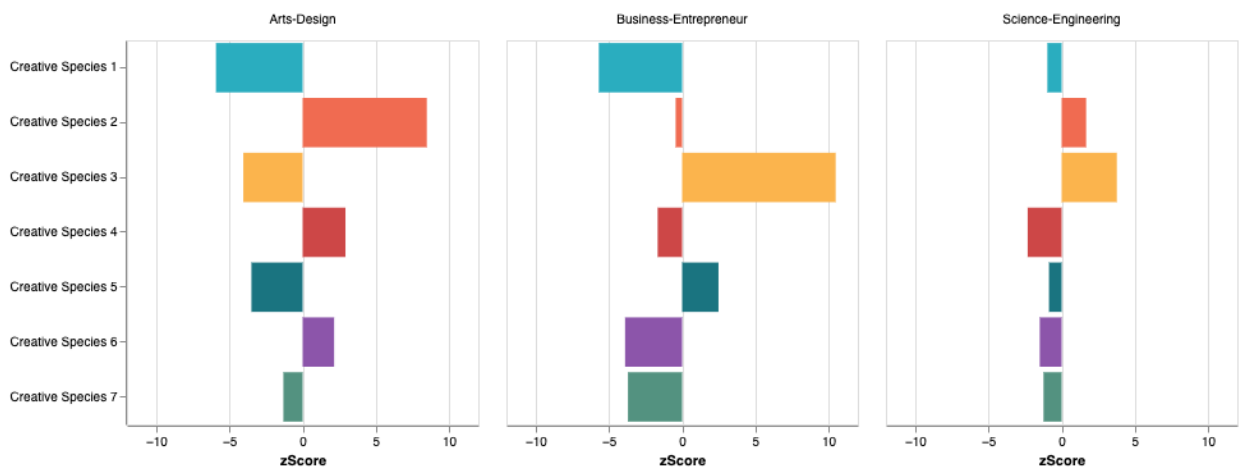

Figure 5. Over- vs under-representation of Creative Species by self-described Gender. Colors are the same as in Figure 1b, and the defining creative habits of each Creative Species cluster can be found in Table 2. a) The percent difference between the observed and permuted average fraction of people in each gender per Creative Species cluster. b) The z-score (number of standard deviations from the permuted average) for the observed fraction of people of each gender category per Creative Species cluster. Both the percent difference and z-scores for each cluster and gender were calculated relative the mean fraction of people of that gender in 1000 random permutations of the gender labels. The large percent difference between Creative Species clusters for "Non-Binary/Non-Conforming" is an artifact of low sample size (0.05% of respondents), and is reflected in the small permuted z-score values.

a)
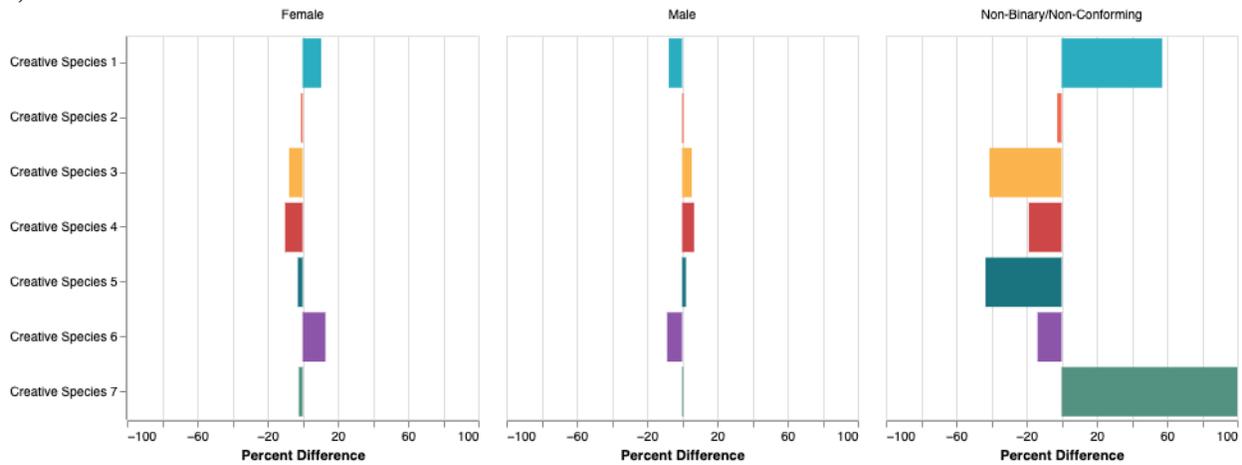

b)
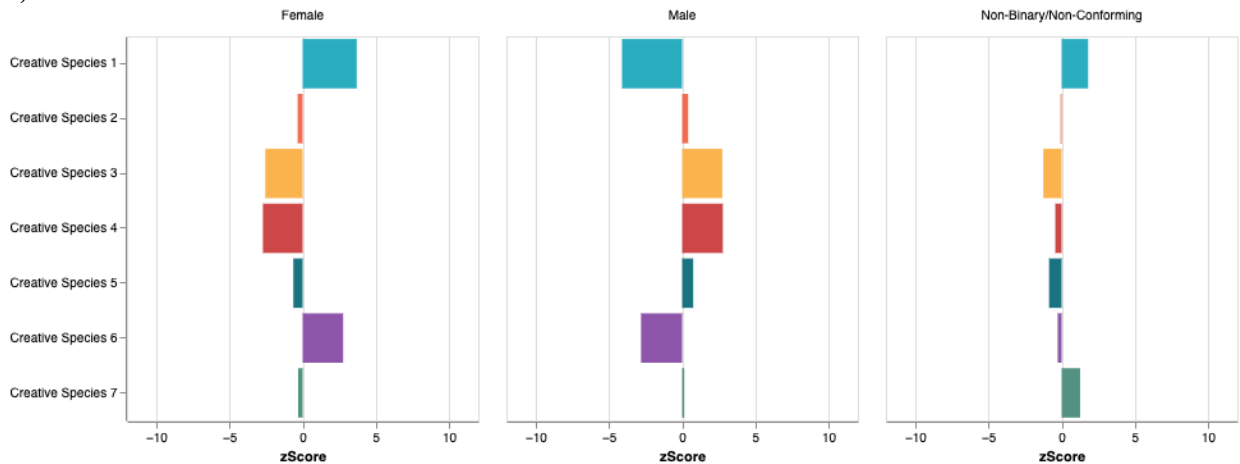

Tables.

Table 1. Summary of the Creative Habit survey used in this study.

| id | Question | Response Scale | Creative Habit | Dimension |
|---|---|---|---|---|
| 1 | On average, how many creative projects do you juggle simultaneously? | I prefer one project at a time:1<br>I prefer many projects at once:5 | **Monotasker**<br>**Multitasker** | Multitasking |
| 2 | What is the breadth of your creative interests? | I specialize in one area:1<br>I work in many areas:5 | **Specialist**<br>**Generalist** | Breadth |
| 3 | How often do you engage in creative collaboration? | I tend to work on solo endeavors:1<br>I often engage in collaborative endeavors:5 | **Solo Creator**<br>**Collaborator** | Collaboration |
| 4 | Do you tend be more self-critical or more self-assured in your creative work? | Self-Critical:1<br>Self-Assured:5 | **Self Critical**<br>**Self Assured** | Confidence |
| 5 | Broadly speaking, how do you feel about the role of distractions in your creative process? | Love Distractions:1<br>Hate Distractions:5 | **Like Distractions**<br>**Dislike Distractions** | Distractions |
| 6 | What do you regard as your strongest source of creative inspiration? | From within (e.g., dreams, meditation, daydreaming):1<br>From outside (e.g., reading, events, conversation):5 | **Inwardly Inspired**<br>**Outwardly Inspired** | Inspiration Source |
| 7 | How do you envision and implement your creative projects? | I know what I want to achieve and execute a plan:1<br>I follow my intuition and let the vision crystalize as I work:5 | **Rational**<br>**Intuitive** | Creative Reasoning |
| 8 | What factors are most significant in motivating your creative work? | Internal (e.g., a need to express myself):1<br>External (e.g., sense of acheivement, pleasing others, earning a living):5 | **Internally Motivated**<br>**Externally Motivated** | Motivation |
| 9 | How important do you think physical exercise and/or movement is to being creative in your work? | Unimportant:1<br>Very important:5 | **NonKinetic**<br>**Kinetic** | Movement |
| 10 | How compelled would you be to clean a messy workspace before beginning your creative work? | Rarely, if ever:1<br>Always:5 | **Comfroting Mess**<br>**Tidy Workspace** | Organization |
| 11 | Compared to others, how rapidly do you tend to generate new ideas or possible solutions? | I need time to let my ideas surface:1<br>I get new ideas by the minute:5 | **Slow-Paced**<br>**Fast-Paced** | Pace |
| 12 | How easy is it for you to do mediocre work if it seems prudent? | It's easy:1<br>It's extremely difficult:5 | **Pragmatist**<br>**Perfectionist** | Perfectionism |
| 13 | When a significant risk is involved in my creative endeavors… | I do what I can to minimize it:1<br>I get excited about the possibiities:5 | **Risk-Averse**<br>**Risk-Friendly** | Risk |
| 14 | To what degree does your creative work involve the element of chance? | Rarely, I make circumstances happen:1<br>Always, I let circumstances happen:5 | **Make It Happen**<br>**Let It Unfold** | Strategy |
| 15 | When your initial plans for creative work don't pan out, how quickly do you typically move to Plan B? | I figure out a way to make Plan A work:1<br>I quickly reframe and move to Plan B:5 | **Tenacious**<br>**Reframer** | Tactic |
| 16 | What kind of space is most productive for you when you are working on a creative project? | Private spaces (e.g., at home, a private office or studio):1<br>Public spaces (e.g., coffee shops, open workspaces):5 | **Private Spaces**<br>**Public Spaces** | Workspace |
| 17 | What noise level is most comfortable for you when you are working on a creative project? | Total silence or minimal background noise:1<br>Loud music or lots of background noise:5 | **Quiet/Silence**<br>**Noise/Loud Music** | Noise/Music |
| 18 | How important to your creative process is it that you spend time outdoors in nature? | Unimportant:1<br>Extremely important:5 | **Natre-Agnostic**<br>**Nature-Lover** | Nature |
| 19 | I'm most creative if my surroundings and routine are… | Varied and dynamic:1<br>Consistent and stable:5 | **Novelty-Seeker**<br>**Routine-Seeker** | Routine |
| 20 | How do you feel about the role of rules in your creative process? | Rules/Constraints stifle my creativity:1<br>Rules/Constraints stimulate my creativity:5 | **Stifled by Constraints**<br>**Stimulated by Constraints** | Rules/Constraints |
| 21 | What time of day do you tend to be most creatively productive? | "Pre Dawn" or "Morning"<br>"Evening" or "After Midnight" | **Early Bird**<br>**Night Owl** | Creative Biorhythm |

Table 2. Defining creative habits of each Creative Species cluster. Creative Species id numbers are the same in all figures. Creative Species are sorted by size in descending order, and the top tags in each are sorted by their 'Weighed Frequency' (Frequency x Sqrt(Relative_Frequency)). See Table 1 for more details about each Creative Habit label. See Box 1 for more detailed descriptions of each Creative Species.

| Cluster | Top Creative Habits | Weight | Frequency | Relative Frequency | Global Frequency | Creative Species |
|---|---|---|---|---|---|---|
| Creative Species 1 | Risk-Averse | 97.85 | 68.34 | 2.05 | 33.27 | Mono routinus |
| | Routine-Seeker | 94.21 | 64.1 | 2.16 | 29.61 | |
| | Dislike Distractions | 83.35 | 69.22 | 1.45 | 47.74 | |
| | Monotasker | 82.75 | 62.73 | 1.74 | 36.03 | |
| | Make It Happen | 78.55 | 56.54 | 1.93 | 29.37 | |
| | Solo Creator | 77.57 | 65.56 | 1.4 | 46.83 | |
| | Private Spaces | 68.99 | 60.05 | 1.32 | 45.33 | |
| | Externally Motivated | 65.71 | 52.44 | 1.57 | 33.47 | |
| | Rational | 64.61 | 47.76 | 1.83 | 26.13 | |
| | Slow-Paced | 61.29 | 46.2 | 1.76 | 26.2 | |
| | Stimulated By Constraints | 60.89 | 48.44 | 1.58 | 30.57 | |
| | Nature-Agnostic | 49.43 | 37.8 | 1.71 | 22.06 | |
| | Specialist | 41.67 | 30.15 | 1.91 | 15.75 | |
| Creative Species 2 | Novelty-Seeker | 87.48 | 71.67 | 1.49 | 48.23 | Yolo chaotis |
| | Risk-Friendly | 84.14 | 68.02 | 1.53 | 44.46 | |
| | Stifled By Constraints | 82.15 | 66.2 | 1.54 | 42.89 | |
| | Internally Motivated | 80.92 | 64.38 | 1.58 | 40.76 | |
| | Multitasker | 77.03 | 60.52 | 1.62 | 37.44 | |
| | Let It Happen | 72.19 | 58.75 | 1.51 | 38.98 | |
| | Inwardly Inspired | 59.86 | 46.46 | 1.66 | 28.04 | |
| | Comforting Mess | 58.47 | 45.52 | 1.65 | 27.51 | |
| | Intuitive | 53.83 | 39.9 | 1.82 | 21.94 | |
| | Noise/Music | 52.05 | 42.5 | 1.5 | 28.25 | |
| | Fast-Paced | 48.77 | 37.97 | 1.65 | 22.95 | |
| Creative Species 3 | Collaborator | 96.21 | 65.01 | 2.19 | 29.73 | Socialis adventurous |
| | Risk-Friendly | 89.15 | 70.7 | 1.59 | 44.46 | |
| | Multitasker | 88.33 | 66.39 | 1.77 | 37.44 | |
| | Reframer | 73.78 | 62.58 | 1.39 | 44.89 | |
| | Kinetic | 72.39 | 56.02 | 1.67 | 33.55 | |
| | Fast-Paced | 44.38 | 35.65 | 1.55 | 22.95 | |
| | Early Bird | 42.88 | 32.89 | 1.7 | 19.34 | |
| | Self-Assured | 39.23 | 29.91 | 1.72 | 17.4 | |
| | Public Spaces | 28.30 | 22.3 | 1.61 | 13.86 | |
| Creative Species 4 | Novelty-Seeker | 82.72 | 69.17 | 1.43 | 48.23 | Focus mononovous |
| | Monotasker | 79.39 | 61.07 | 1.69 | 36.03 | |
| | Stifled By Constraints | 72.42 | 60.77 | 1.42 | 42.89 | |
| | Dislike Distractions | 68.16 | 60.48 | 1.27 | 47.74 | |
| | Nature-Lover | 58.51 | 45.69 | 1.64 | 27.88 | |
| Creative Species 5 | Novelty-Seeker | 91.11 | 73.66 | 1.53 | 48.23 | Novo gregarious |
| | Externally Motivated | 81.38 | 60.49 | 1.81 | 33.47 | |
| | Outwardly Inspired | 79.35 | 63.53 | 1.56 | 40.63 | |
| | Collaborator | 71.30 | 53.29 | 1.79 | 29.73 | |
| | Noise/Music | 69.44 | 51.47 | 1.82 | 28.25 | |
| | Stimulated By Constraints | 64.95 | 50.56 | 1.65 | 30.57 | |
| | Love Distractions | 57.81 | 42.05 | 1.89 | 22.19 | |
| | Public Spaces | 52.60 | 33.74 | 2.43 | 13.86 | |
| Creative Species 6 | Routine-Seeker | 82.70 | 58.77 | 1.98 | 29.61 | Sui inspira |
| | Private Spaces | 79.17 | 65.75 | 1.45 | 45.33 | |
| | Internally Motivated | 78.70 | 63.21 | 1.55 | 40.76 | |
| | Solo Creator | 74.09 | 63.53 | 1.36 | 46.83 | |
| | Inwardly Inspired | 64.42 | 48.84 | 1.74 | 28.04 | |
| Creative Species 7 | Stifled By Constraints | 77.34 | 63.57 | 1.48 | 42.89 | Solo noctus |
| | Internally Motivated | 74.02 | 60.64 | 1.49 | 40.76 | |
| | Solo Creator | 73.29 | 63.08 | 1.35 | 46.83 | |
| | Night Owl | 70.92 | 60.15 | 1.39 | 43.32 | |
| | Risk-Averse | 69.82 | 54.52 | 1.64 | 33.27 | |
| | Make It Happen | 64.02 | 49.39 | 1.68 | 29.37 | |
| | Tenacious | 63.38 | 48.9 | 1.68 | 29.03 | |
| | NonKinetic | 54.96 | 39.36 | 1.95 | 20.16 | |
| | Nature-Agnostic | 44.54 | 35.21 | 1.6 | 22.06 | |
| | Love Distractions | 40.72 | 33.25 | 1.5 | 22.19 | |
| | Specialist | 38.05 | 28.36 | 1.8 | 15.75 | |

SUPPLEMENTARY INFORMATION

SI Table 1. Questions from the first, exploratory version of the survey with 270 questions which was tested on 1000 people crowd-sourced from a micro-work platform.

See here: https://www.dropbox.com/s/f6pgvsyapjosonn/SI-Table1-CreativeStyle-Survey-Phase1-CrowdFlower.pdf?dl=0

SI Table 1. Questions from the second version of the survey with 91 questions which was tested on 500 accomplished people in Red Bull's High Performance community.

See here: https://www.dropbox.com/s/1nc1r8qbd38vz6q/S1-Table_2_CreativeStyle-Survey-Phase2-HighPerformers.pdf?dl=0

SI Figure 1. Stability of the Creative Species clusters (groups of people that tend to share similar combinations of creative habits). Each point is a Creative Species cluster from the network of people linked by similar sets of creative habits, where the target average links per node was 10. The clusters are sized, and vertically sorted in descending order, by the fraction of people in that cluster. The x-axis is the average (+/- 1 standard deviation) weighted Jaccardian similarity of each cluster to the most similar cluster in a) 1000 networks generated from random samples of 50% of the population, and b) different target link densities in the network generation algorithm (6, 7, 8, 9, 11, 12). The grey points were unstable, small clusters which were all <1% of the population. Colored clusters are the seven largest and most stable, and colors are the same as in the figures, and of the main text. The defining creative habits of each stable Creative Species cluster can be found in Table 2 of the main text.

a)

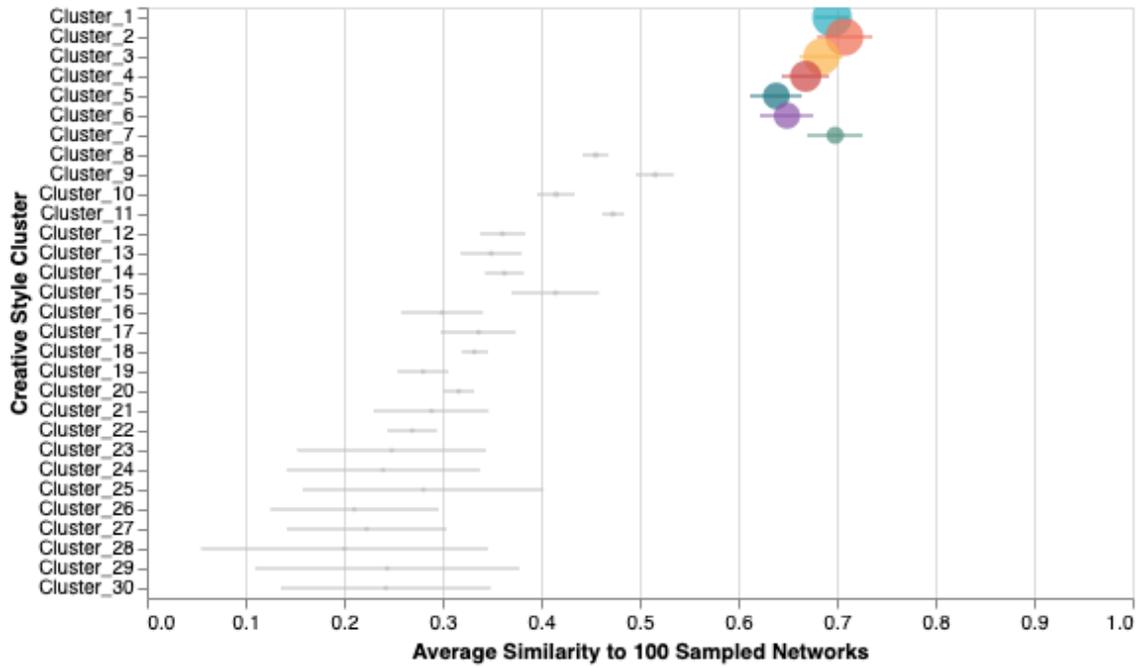

b)

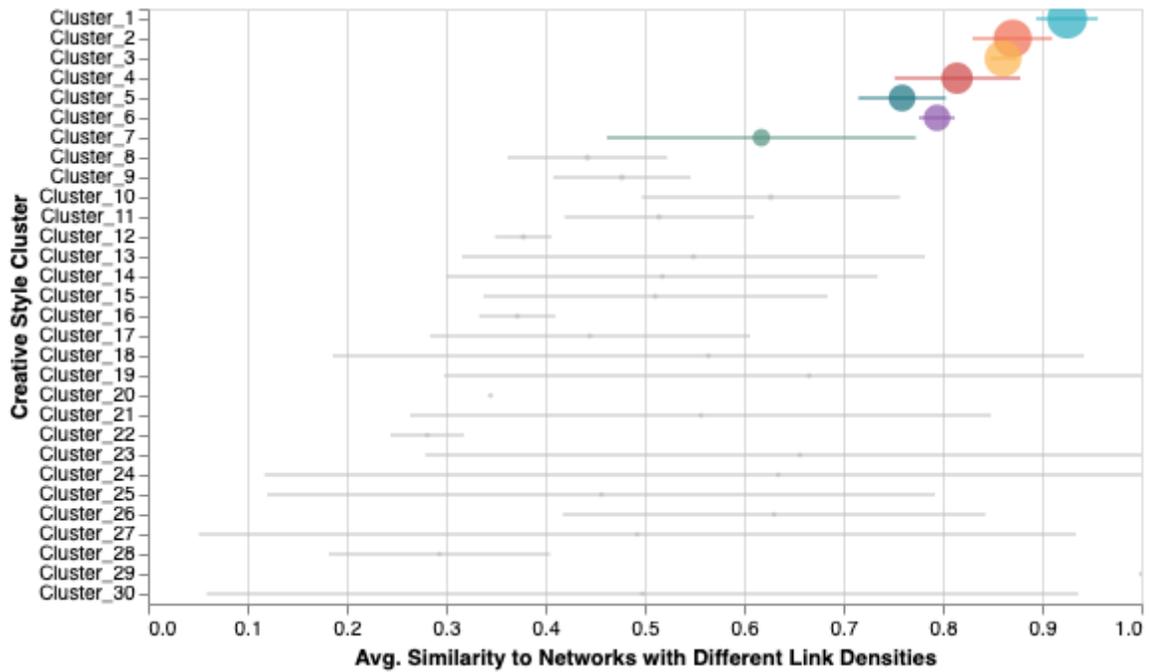